# Machine learning meets network science: dimensionality reduction for fast and efficient embedding of networks in the hyperbolic space

Josephine Maria Thomas[1,†], Alessandro Muscoloni[1,2,†], Sara Ciucci[1,3], Ginestra Bianconi[4] and Carlo Vittorio Cannistraci [1,*]

[1]Biomedical Cybernetics Group, Biotechnology Center (BIOTEC), Center for Molecular and Cellular Bioengineering (CMCB), Technische Universität Dresden, Tatzberg 47/49, 01307 Dresden, Germany

[2]School of Science, ALMA MATER STUDIORUM - Università di Bologna - Via Zamboni, 33 - 40126 Bologna, Italy

[3]Lipotype GmbH, Tatzberg 47, 01307 Dresden, Germany

[4]School of Mathematical Sciences, Queen Mary University of London, London E1 4NS, United Kingdom

*Correspondence should be addressed to: kalokagathos.agon@gmail.com

†The first two authors should be regarded as joint First Authors

**Complex network topologies and hyperbolic geometry seem specularly connected (Papadopoulos et al. 2012), and one of the most fascinating and challenging problems of recent complex network theory is to map a given network to its hyperbolic space. The Popularity Similarity Optimization (PSO) model represents - at the moment - the climax of this theory (Papadopoulos et al. 2012). It suggests that the trade-off between node popularity and similarity is a mechanism to explain how complex network topologies emerge - as discrete samples - from the continuous world of hyperbolic geometry (Papadopoulos et al. 2015). The hyperbolic space seems appropriate to represent real complex networks. In fact, it preserves many of their fundamental topological properties, and can be exploited for real applications such as, among others, link prediction and community detection. Here, we observe for the first time that a topological-based machine learning class of algorithms -** *for nonlinear unsupervised dimensionality reduction* **- can directly approximate the network's node angular coordinates of the hyperbolic model into a two-dimensional space, according to a similar topological organization that we named** *angular coalescence.* **On the basis of this phenomenon, we propose a new class of algorithms that offers fast and accurate** *coalescent embedding* **of networks in the hyperbolic space even for graphs with thousands of nodes.**

Significant progress has been achieved in the last fifteen years in unveiling the universal properties of complex networks. Nevertheless, characterizing the large variety of real network structures which are coming from the 'Big Data explosion' remains an important challenge of network science. *Network geometry* aims at making a paradigmatic shift in our understanding of complex network structures by revealing their hidden metric (Boguñá et al. 2008; Papadopoulos et al. 2012; Higham et al. 2008; Kuchaiev et al. 2009; Wu et al. 2015; Bianconi & Rahmede 2015; Cannistraci et al. 2013a; Cannistraci et al. 2013b; Daminelli et al. 2015). This field has a large number of applications ranging from brain networks (Cannistraci et al. 2013a) to routing packets in the Internet (Bianconi 2015). In this context there is increasing evidence that the hidden metric of many complex networks is hyperbolic. However, the science that studies and designs algorithms to test this hypothesis and to reveal the hidden hyperbolic metric of real complex networks, is in its dawning. To this purpose, here we propose advanced algorithms that combine network geometry, manifold theory and machine learning with the PSO model.

The PSO model suggests that real networks have a congruous geometrical representation in a hyperbolic space, where each network node is mapped according to the angular and the radial coordinates of a polar system (Papadopoulos et al. 2012). The node similarities are related with the angular distances in the hyperbolic space: the higher the similarity between two nodes, the closer their angular coordinates. On the other hand, the node degree is related with the intrinsic popularity of the node: the higher the node degree, the higher its popularity in the network and the lower its radial coordinate in the hyperbolic space.

Manifold machine learning for unsupervised nonlinear dimensionality reduction are an important sub-class of topological machine learning. They learn nonlinear similarities (or distances) between points (samples) distributed over a hidden manifold in a multidimensional feature space, in order to preserve, embed (map) and visualize them in a two-dimensional reduced space (Cannistraci et al. 2010). They are inspired by a three-step procedure. First, they approximate the shape of the hidden manifold reconstructing a nearest-neighbourhood graph between the points in the high-dimensional space. Second, they use the reconstructed network to estimate pairwise topological similarities (or distances) between the points that lie on the manifold, and store these nonlinear estimations in a kernel (or distance matrix). In a third and last step, they apply a matrix decomposition to the kernel to perform dimensionality reduction, usually in a space of two-dimensions. If the network is already given in the form of an unweighted adjacency matrix, the same algorithm works neglecting the first step and thus, in practice, performs a network embedding that preserves the node similarities. These methods are already used in network biology for instance to predict node similarities in protein interaction networks (Kuchaiev et al. 2009; Cannistraci et al. 2013b), therefore it was likely for us to envisage their usage for network embedding in the hyperbolic space.

In Fig. 1B-D, we show that the Isomap algorithm (ISO), which is the progenitor of manifold dimension reduction techniques, starting from the unweighted adjacency matrix offers an embedding of the network nodes that is organized according to a circular pattern (Fig. 1B), that follows the angular coordinates of the original PSO model. Unexpectedly, for an algorithm named noncentered minimum curvilinear embedding (ncMCE) (Cannistraci et al. 2013b) (Fig. 1E-F), the circular pattern is linearized (Fig. 1E) and the nodes are ordered (respecting the node similarities) on the second dimension of embedding. In fact, if we accommodate the node points on the circumference following the same ordering of the second dimension of embedding (Fig. 1E) we can again recover an unbroken circular pattern (Fig. 1F) that respects the angular coordinates of the original PSO model. Surprisingly, ncMCE is the only algorithm - between the ones tested - able to learn, unfold and linearize along just one dimension an intrinsic nonlinear (circular) pattern. It compresses the information in one unique dimension because it learns nonlinear similarities by means of the minimum spanning tree. In fact, ncMCE provides a *hierarchical-based* mapping that is fundamentally different from the *manifold-based* of ISO. The capability to embed nonlinear variability information exclusively in one dimension is highly valued in machine learning because demonstrates the power of dimension reduction techniques to linearize nonlinear patterns (Cannistraci et al. 2010; Roweis & Saul 2000) (see Methods). However, the utility of a computational technique varies in relation to the designed target to reach. ncMCE is based on a general nonlinear similarity learning theory called Minimum Curvilinearity (Cannistraci et al. 2010; Cannistraci et al. 2013b) that offers an advantage over manifold approaches when the purpose is unsupervised pattern detection and linearization. In fact, in that case the goal is to perform embedding of the points of a multidimensional datasets, and the graph that connects the points and approximates the manifold is unknown (see Methods). Conversely, when the task is to embed a network in the hyperbolic space, the graph that approximates the manifold is already given, and the goal of embedding is to retain - and not to linearize - the circular node similarity pattern in the two-dimensional space. Therefore the manifold-based techniques, compared to the minimum-curvilinear-based, should offer a better approximation of the angular coordinates which represent the network node similarities; especially for high temperatures of the PSO model, when the tree-like network organization and the related hyperbolic geometry degenerate.

The rationale of our approach is all contained here. We tried many combinations of different network similarities and matrix decompositions. And, in brief, we reached the same consistent finding: the arising in the two-dimensional embedding space of a common node aggregation pattern which we named *angular coalescence*, and that was circularly or linearly ordered according to the angular coordinates of the hyperbolic model. This represents the first important discovery of our study. Consequently, we decided to coin the expression *coalescent embedding* to indicate the class of algorithms that exhibit angular coalescence in the two-dimensional network embedding. In our case we detected the angular coalescence phenomenon as embedding result of topological-based machine learning for nonlinear unsupervised dimension reduction. Indeed, the evidence that even MCE and ncMCE are able to exhibit coalescent embedding may theoretically suggest that this is an *epiphenomenon* that in general characterizes topological-based machine learning for nonlinear dimension reduction when applied to this task.

In the end, for this study we decided to select a representative group of nonlinear unsupervised dimensionality reduction approaches among the ones with the highest performance. Three manifold-based: Isomap (ISO) (Tenenbaum et al. 2000), noncentered Isomap (ncISO) (Cannistraci et al. 2013b), Laplacian eigenmaps (LE) (Belkin & Niyogi 2001). Two minimum-curvilinearity-based: minimum curvilinear embedding (MCE) (Cannistraci et al. 2010; Cannistraci et al. 2013b) and noncentered minimum curvilinear embedding (ncMCE) (Cannistraci et al. 2013b). An important note for practical applications is that all these approaches are unsupervised (node or edge labels are not required) and parameter-free (external tuning setting of algorithms' parameters is not required).

In Fig. 2, we propose a novel general algorithm - based on the coalescent embedding principle - for network embedding in the hyperbolic space. In order to build a general algorithm we started by noticing that the problem to compute the embedding on an unweighted adjacency matrix would be simplified by having a 'good guess' of the edge weights that suggest the connectivity geometry. Thus, there was a clear margin to improve the coalescent embedding performance by *pre-weighting* the network links using a convenient strategy to suggest topological similarities between the connected nodes. We devised two different pre-weighting strategies. The first adopts what we called the repulsion-attraction rule (RA). The idea is that links that are between nodes with high degree should be geometrically far because they represent hubs, and big hubs - according to the theory of navigability of complex networks presented by Boguñá et al. (Boguñá et al. 2008) - tend to dominate geometrically distant regions: this is the repulsive part of the rule. On the contrary, links between nodes that share a high number of common neighbours should be geometrically close because they most likely share many similarities: this is the attractive part of the rule. Thus, the RA (see Fig. 2 for the precise mathematical formula) is a simple and efficient approach that quantifies the trade-off between *hub repulsion* and *common-neighbours-based attraction*. It pre-weights the network links adding up the adjacent nodes' degrees and their cross term (the product), and then dividing by the number of common neighbours. Suppl. Fig. 1 is giving a visual example of how the RA pre-weighting rule is improving the angular coalescence effect in respect to the same methods adopted without pre-weighting in Fig. 1. Inspired by the same rationale, the second strategy makes a similar pre-weighting of the links using the edge-betweenness-centrality (EBC) to obtain geometrical distances between nodes and regions of the networks. Furthermore, we were not convinced that preserving the final angular distances between adjacent nodes (Fig. 2B) was the best strategy to re-organize the nodes over the disk circumference. Most likely the reciprocal angular distances between adjacent nodes on the circumference were affected by *local noise*. Thus we devised a new strategy to re-organize the nodes on the circumference (Fig. 2B) that we called equidistant adjustment (EA): the nodes are equidistantly re-organized on the circumference according to their original order learned by the coalescent embedding. Fig. 2A-C offers an example of the process of equidistant adjustment for the LE embedded nodes.

The results for the best dimension reduction methods are reported in Fig. 3A. The performance was evaluated as Pearson correlation between all the pairwise hyperbolic distances of the network nodes in the original PSO model and in the reconstructed hyperbolic space. The plots report the average correlation over the 100 synthetic networks that have been generated for each different PSO model parameter combination. It is evident that the coalescent embedding techniques pre-weighted with RA and adjusted according to EA are outperforming the Hypermap algorithm that is the state of the art, and this is the second key discovery of our study. RA performed similarly to EBC, and in general both the pre-weighting strategies are effective (Suppl. Fig. 2-6). However, RA is computational more efficient. Obviously, all the methods reduce their performance for increasing temperature (reduced clustering), because the networks assume a more random structure.

Another alluring result, pointing out a very subtle problem, is that without EA all techniques significantly reduce the performance, as it is shown in Fig. 3B. Looking at Fig. 3A-B and the Suppl. Fig. 2-6, EA makes a difference especially for low temperatures (high clustering), while for high temperatures its improvement is vanishing. This is particularly evident for LE that in Fig. 3B at low temperatures has a significantly worse performance compared to Fig.

3A where EA is applied. Imposing an equidistant adjustment might be counterintuitive, but our simulations suggest that this sub-optimal strategy is better than passively undergo the *local embedding uncertainty*. On the other hand, once the temperature is increased, the *global embedding uncertainty* also increases and the techniques are less efficient to recover the global node order. In practice, for high temperatures the global noise overcomes the local noise and the EA reduces its effectiveness. The significant improvement offered by EA is evident also in link prediction simulations on real world networks (see Table 1), where EA-based coalescent embedding techniques offers the best precision, confirming that adjustment of local embedding uncertainty is crucial for effective coalescent embedding also in real applications. Taken all together our results suggest that the idea to connect manifold learning theory with network geometry is a promising direction of research. In fact, the coalescent-embedding-based algorithms combine important performance improvement with a spectacular speed up - both on *in-silico* and real tests - in respect to Hypermap (see Fig. 4 and Table 1).

To conclude, we hope that this letter will contribute to establish a new bridge at the interphase between physics of complex networks and computational machine learning theory, and that future extended studies will dig into real applications revealing the impact of coalescent network embedding for instance in network medicine or social science predictions.

**Methods**

**Generation of synthetic networks by the PSO model**

The synthetic networks used in the simulations have been created according to the PSO model (Papadopoulos et al. 2012), which describes how random geometric graphs grow in the hyperbolic space. The model has four input parameters: $m > 0$, which defines the average node degree $\bar{k} = 2m$, $\beta \in (0,1]$, defining the exponent $\gamma = 1 + 1/\beta$ of the power law degree distribution, $T \geq 0$, which controls the network clustering, and $\zeta = \sqrt{-K} > 0$, where $K$ is the curvature of the hyperbolic plane. The network clustering is maximized at $T = 0$, it decreases almost linearly for $T = [0,1)$ and it becomes asymptotically zero if $T > 1$.

Building a network of $N$ nodes on a hyperbolic plane with curvature $K = -1$ requires the following steps: (1) Initially the network is empty; (2) At time $i = 1, 2, \ldots, N$ a new node $i$ appears with radial coordinate $r_i = 2 \ln(i)$ and angular coordinate $\theta_i$ uniformly sampled from $[0, 2\pi]$; all existing nodes $j < i$ increase their radial coordinates according to $r_j(i) = \beta r_j + (1 - \beta) r_i$ in order to simulate popularity fading; (3) The new node picks a randomly chosen existing node and connects to it with probability $p(h_{ij}) = 1/(1 + exp((h_{ij} - R_i)/2T))$, where $R_i = r_i - 2\ln\left[\frac{2T(1-e^{-(1-\beta)\ln(i)})}{\sin(T\pi)m(1-\beta)}\right]$ is the current radius of the hyperbolic disk, $h_{ij} = \text{arccosh}(\cosh r_i \cosh r_j - \sinh r_i \sinh r_j \cos\theta_{ij})$ is the hyperbolic distance between node $i$ and node $j$ and $\theta_{ij} = \pi - |\pi - |\theta_i - \theta_j||$ is the angle between these nodes. Step (3) is repeated until the new node $i$ is connected to $m$ nodes. (4) The growing process stops when $N$ nodes have been introduced.

In this model networks evolve optimizing a trade-off between node popularity, abstracted by the radial coordinate, and similarity, represented by the angular coordinate, and they exhibit many common structural and dynamical characteristics of real networks.

**HyperMap**

HyperMap (Papadopoulos et al. 2015) is a method based on Maximum Likelihood Estimation to map a network into its hyperbolic space. It replays the hyperbolic growth of the network and at each step it finds the polar coordinates of the added node by maximizing the likelihood that the network was produced by the E-PSO model (Papadopoulos et al. 2015).

For curvature $K = -1$ the procedure is as follows: (1) Nodes are sorted decreasingly by degree and then labeled $i = 1, 2, ..., N$ according to the order; (2) Node $i = 1$ is born and assigned radial coordinate $r_1 = 0$ and a random angular coordinate $\theta_1 \in [0, 2\pi]$; (3) For each node $i = 2, 3, ..., N$ do: (3.a) Node $i$ is added to the network and assigned a radial coordinate $r_i = 2\ln(i)$; (3.b) The radial coordinate of every existing node $j < i$ is increased according to $r_j(i) = \beta r_j + (1-\beta)r_i$; (3.c) The node $i$ is assigned an angular coordinate by maximizing the likelihood $L_i = \prod_{1 \leq j < i} p(h_{ij})^{x_{ij}}(1 - p(h_{ij}))^{1-x_{ij}}$, where $\beta$ and $p(h_{ij})$ are the same as in the PSO model and $x_{ij}$ is the adjacency matrix. The maximization is done by numerically trying different angular coordinates in steps of $2\pi/N$ and choosing the one that leads to the biggest $L_i$.

**Radial coordinates**

A Maximum Likelihood Estimation of the sequence according to which nodes appeared in the networks indicates that the higher the degree of the node, the earlier it appeared (Papadopoulos et al. 2015). Therefore, nodes are sorted by descending degree and then labeled $i = 1, 2, ..., N$ according to this order. Considering $K = -1$, the radial coordinates are computed as follows (Wang et al. 2016): $r_i = 2(\beta \ln i + (1-\beta) \ln N)$. The parameter $\beta$ is computed as $\beta = 1/(\gamma - 1)$. The exponent $\gamma$ of the power law degree distribution has been fitted using the MATLAB script *plfit.m*, an algorithm published at

**Manifold-based embedding**

The first type of topological-based unsupervised machine learning for nonlinear dimension reduction adopted in this study are Isomap (ISO) (Tenenbaum et al. 2000) and Laplacian Eigenmaps (LE) (Belkin & Niyogi 2001). These two methods are manifold-based machine learning because, in classical dimension reduction of multidimensional datasets, they approximate the sample data manifold using a proximity graph, and then they embed the sample distances in a two-dimensional space by matrix decomposition. In our application the proximity graph is already given, representing an important advantage, because the topological connections (similarities) between the nodes are already known. In fact, the problem to infer a proximity graph is not trivial and generally requires the introduction of at least a tuning parameter, for instance in the procedure to learn a nearest-neighbour graph (network) that approximate the manifold. Furthermore, there is not a clear strategy to unsupervisedly tune these kinds of parameters to infer the proximity graph.

ISO is based on extracting a distance matrix (or kernel) that stores all the network shortest path distances (also named geodesic distances) that approximate the real distances over the manifold. Then the kernel is centred and in this work singular value decomposition (SVD) is applied to embed the nodes in the two-dimensional space. We also propose the noncentered version of the same algorithm, named ncISO, in which the kernel centering is neglected. Consequently, the first dimension of embedding is discharged because, since it points toward the center of the manifold, is not useful. For more computational details on the implementation of ISO please refer to (Tenenbaum et al. 2000; Cannistraci et al. 2013b).

LE is a different type of manifold machine learning. In fact, the inference of a distance kernel (for instance the shortest path kernel for ISO) starting from the network structure is not required in this algorithm, which makes it faster than ISO. Indeed, the idea behind LE is to perform the eigen-decomposition of the network's Laplacian matrix, and then to perform two-dimensional embedding of the network's nodes according to the eigenvectors related to the second and third smallest eigenvalues. The first smallest eigenvalue is zero, thus the related eigenvector is neglected. In order to implement a weighted version of this algorithm we used, as suggested in the original publication (Belkin & Niyogi 2001), the 'heat-function' (instead of the pre-weighting values as they are in their original scale):

$$\tilde{x}_{ij} = e^{-\frac{x_{ij}^2}{t}}$$

Where $x_{ij}$ is the original pre-weighing value for the link *i,j*, and *t* is a scaling factor fixed as the mean of all the network's pre-weighting values. For more computational details on the implementation of LE please refer to (Belkin & Niyogi 2001).

**Minumum Curvilinearity and Minimum curvilinear embedding**

The centered and noncentered versions of the minimum curvilinear embedding algorithm – respectively named MCE and ncMCE - are based on a general nonlinear similarity learning theory called Minimum Curvilinearity (Cannistraci et al. 2010; Cannistraci et al. 2013b). These approaches compress the information in one unique dimension because they learn nonlinear similarities by means of the minimum spanning tree, providing a hierarchical-based mapping. This is fundamentally different from the previous algorithms (ISO and LE), which are manifold-based. If we would consider the mere unsupervised machine learning standpoint, we would notice that manifold-based techniques in this study showed two main weaknesses: i) they offer less compression power because two orthogonal dimensions of representation, instead of one, are needed; ii) the node similarity pattern remains nonlinear (circular) also in the embedded space, thus the goal of the nonlinear dimension reduction to linearize a (hidden) nonlinear pattern in the embedding space is missed. In unsupervised tasks where the objective is to discover unknown and unexpected sample stratifications - for instance the discovery of unforeseen groups of patients with undetected molecular-based disease variations – the linearization of a pattern along one unique embedding dimension can offer an undisputed help to recognize hidden and hierarchical organized subgroups (Roweis & Saul 2000; Cannistraci et al. 2010). Interestingly, MCE and ncMCE were theoretically designed according to a previous theory of Boguñá, et al. (Boguñá et al. 2008). When the last author of this article, read at the beginning of 2009 for the first time the article "Navigability of complex networks" (Boguñá et al. 2008), he had an intuition. This article was clearly explaining that to efficiently navigate a network (and thus approximate geodesic/curvilinear pairwise node connections over the hidden manifold) it was not necessary to know the complete information of network topology at the starting point of the navigation. A greedy routing process (thus, based on the neighborhood information) was enough to efficiently navigate the network. This triggered an easy conclusion: to approximate curvilinear distances between the points of the manifold it was not necessary to reconstruct the nearest-neighbour graph. Just a greedy routing process (that exploits a norm, for instance Euclidean) between the points in the multidimensional space, is enough to efficiently navigate the hidden network that approximates the manifold in the multidimensional space. In a few words, learning nonlinear distances over the manifold by navigating an invisible and unknown network was possible, because the navigation process was instead guided by a greedy routing. And a preferable greedy routing strategy, was the minimum spanning tree (MST). The only hypothesis of application of this approach was that the points were not homogenously distributed in a lattice regular structure, or a similar degenerative condition. Thus, the minimum curvilinearity kernel is the matrix that collects all the pairwise distances between the points (or nodes) computed over the MST. And the ncMCE is the embedding of the noncentered minimum curvilinear kernel by means of the SVD. The reason why to exploit the ncMCE - noncentered version of MCE (Cannistraci et al. 2013b) - is discussed in a second article (Cannistraci et al. 2013b) that presents how to use this approach for link prediction in protein interaction networks. The main difference between MCE and ncMCE is that in general MCE linearizes the hidden patterns along the first dimension of embedding while ncMCE along the second dimension (since it is noncentered the first dimension of embedding should be generally neglected because points towards the center of the manifold). To conclude this part, MCE/ncMCE are conceptually different from all the other approaches because they are among the few (maybe the only, to the best of our knowledge) dimensionality reduction methods that performs *hierarchical embedding*, and they exploit the MST as a highway to navigate different regions of the network. Although they exploit a small fraction of the network links – practically only the MST, which consists of N-1 links in a network with N nodes - the reason why they work efficiently to infer the angular coordinates of networks that follow the PSO model is well explained in the article of Papadopoulos et al. (Papadopoulos et al. 2015), thus we take the advantage to report the full paragraph: << This work shows that random geometric graphs in hyperbolic spaces are an adequate model for complex networks. The high-level explanation of this connection is that complex networks exhibit *hierarchical, tree-like organization*, while hyperbolic geometry is the geometry of trees. Graphs representing complex networks appear then as discrete samples from the continuous world of hyperbolic geometry. >>

However, the problem to compute the MST on an unweighted adjacency matrix is that we do not have a norm that suggests the hidden connectivity geometry. Thus, there was a clear margin to improve the performance of MCE/ncMCE by pre-weighting the links in the network (and the adjacency matrix) using a convenient strategy to suggest topological distances between the connected nodes. In fact, in Fig. 3 and Table 1 we notice that the pre-weighting strategy significantly boosts MCE/ncMCE performance.

**Real network datasets**

Information on the real networks used in the link prediction evaluation are provided in Suppl. Info.

**Hardware and software details**

MATLAB code was used for all the methods and simulations, which were carried out on a Dell workstation under Windows 7 professional with 192 GB of RAM and 2 Intel(R) Xenon(R) X5550 processors with 2.66 and 2.67 GHz.

**Acknowledgments**

We thank: Gregorio Alanis Lobato for the useful suggestions; Ewa Aurelia Miendlarzewska for proofreading the article; Ralph Gey and Alexander Mestiashvili for their IT support.

**Funding**

This work was mainly supported by the independent group leader starting grant of the Technische Universität Dresden (TUD). We acknowledge also support by the Centre for Information Services and High Performance Computing (ZIH) of the TUD. AM was partially supported by the Erasmus+ Mobility for traineeships programme of the University of Bologna.

**Author contributions**

CVC envisage the study, invented and designed the coalescent embedding, the pre-weighting and equidistant adjustment strategies, the algorithms and the experiments. GB suggested to adopt the betweeness centrality measures as a pre-weighting option. JMT, AM and SC implemented and ran the codes and performed the computational analysis with CVC help. All the authors analysed the results. CVC designed the figures and wrote the article with the main aid of GB and with input and corrections from all the other authors. JMT, AM and SC realized the figures under the CVC guidance. CVC led, directed and supervised the study.

**Competing interests**

The authors declare no competing financial interests.

**Data and materials availability**

The MATLAB code will be available after article publication at: https://sites.google.com/site/carlovittoriocannistraci/5-datasets-and-matlab-code.

**Figures**

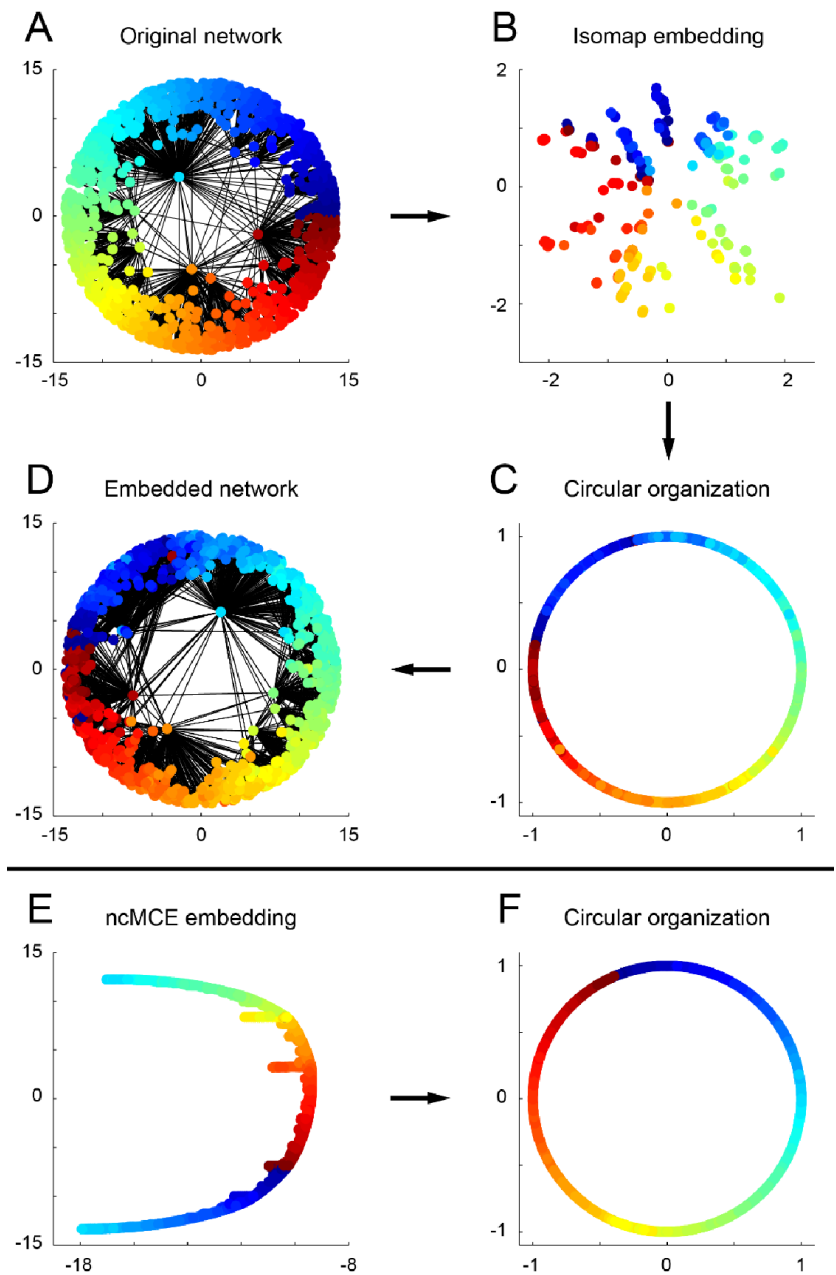

**Fig. 1. Coalescent embedding. (A)** We show the original synthetic network generated by the PSO model in the hyperbolic space. **(B)** Surprisingly, the Isomap algorithm (ISO), which is the progenitor of manifold techniques, starting from the unweighted adjacency matrix offers an embedding of the network nodes that is organized according to a circular pattern that follows the angular coordinates of the original PSO model. We made different trials using other synthetic networks, and this circular pattern is mainly preserved if the kernel is centered or if the kernel is not centered and the first dimension is neglected (see methods for details). This makes sense because the operation of kernel centering puts the origin of the reduced space at the center of the points in a multidimensional space and thus at the center of the manifold. Since the node points lie on the hyperbolic disk, the embedding places the origin approximatively at the center of the disk. **(C)** The nodes are projected over a circumference and adjusted equidistantly according to the step 3.2 of the algorithm described in Fig. 2. **(D)** The radial coordinates are given according to the step 4 of the algorithm described in Fig 2. **(E)** A different pattern is obtained for an algorithm named ncMCE. Unexpectedly, the circular pattern is linearized and the nodes are ordered (respecting the node similarities) on the second dimension of embedding (here the kernel is noncentered and the first dimension of embedding should be neglected, see methods). **(F)** In fact, if we accommodate the node points on the circumference following the same ordering of the second dimension of embedding we can again recover an unbroken circular pattern that respects the angular coordinates of the original PSO model.

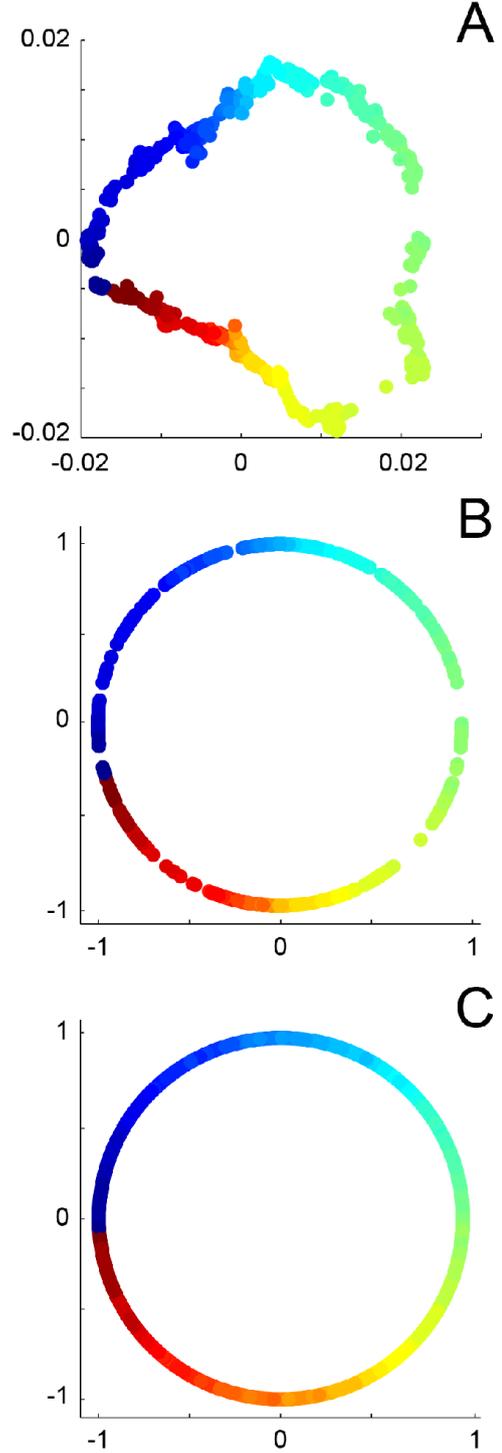

INPUT: adjacency matrix, $x$
OUTPUT: polar coordinates of nodes $(r, \theta)$

1. Pre-weighting rules
   1.1. Repulsion – Attraction Rule (RA)
   $$x_{ij}^{RA} = \begin{cases} \dfrac{d_i + d_j + d_i d_j}{1 + CN_{ij}} & x_{ij} = 1 \\ 0 & x_{ij} = 0 \end{cases}$$
   $x_{ij}$ value of $(i,j)$ entry in matrix $x$
   $d_i$ degree of node $i$
   $CN_{ij}$ common neighbours of nodes $i$ and $j$
   1.2. Edge Betweenness Centrality Rule (EBC)
   $$x_{ij}^{EBC} = \sum_{s,t} \frac{\sigma(s,t|e_{ij})}{\sigma(s,t)}$$
   $V$ set of nodes
   $s,t$ any combination of network nodes in $V$
   $\sigma(s,t)$ number of shortest paths $(s,t)$
   $\sigma(s,t|e_{ij})$ number of shortest paths $(s,t)$ passing through edge $e_{ij}$

2. Nonlinear dimension reduction
   2.1. Laplacian Eigenmaps (LE)
   2.2. Isomap (ISO)
   2.3. Noncentered Isomap (ncISO)
   2.4. Minimum Curvilinearity (MCE)
   2.5. Noncentered Minimum Curvilinearity (ncMCE)

3. Angular Coordinates $(\theta)$
   3.1. Circular adjustment (panel B)
   3.2. Equidistant angular adjustment (panel C)

4. Radial coordinates $(r)$
   Nodes are sorted according to descending degree and the radial coordinate of the ith node in the set is computed according to:
   $$r_i = \frac{2}{\zeta}[\beta \ln i + (1 - \beta) \ln N] \qquad i = 1, 2, ..., N$$
   $N$ number of nodes
   $\zeta = \sqrt{-K}$ (we set $\zeta = 1$)
   $K$ curvature of the hyperbolic space
   $\beta$ popularity fading factor fitted from the network

**Fig. 2. General Algorithm - based on the Coalescent Embedding principle - for network embedding in the hyperbolic space.** For each main step in the algorithm, the sub-steps represent different options to adopt. **(A)** Example of coalescent embedding of a synthetic network by LE. **(B)** Circular adjustment: the nodes are adjusted on a circumference maintaining the original angular coordinates of their coalescent embedding. **(C)** Equidistant angular adjustment: the nodes are adjusted on a circumference maintaining the original angular ordering of their coalescent embedding, but imposing the same distance between adjacent nodes on the ring.

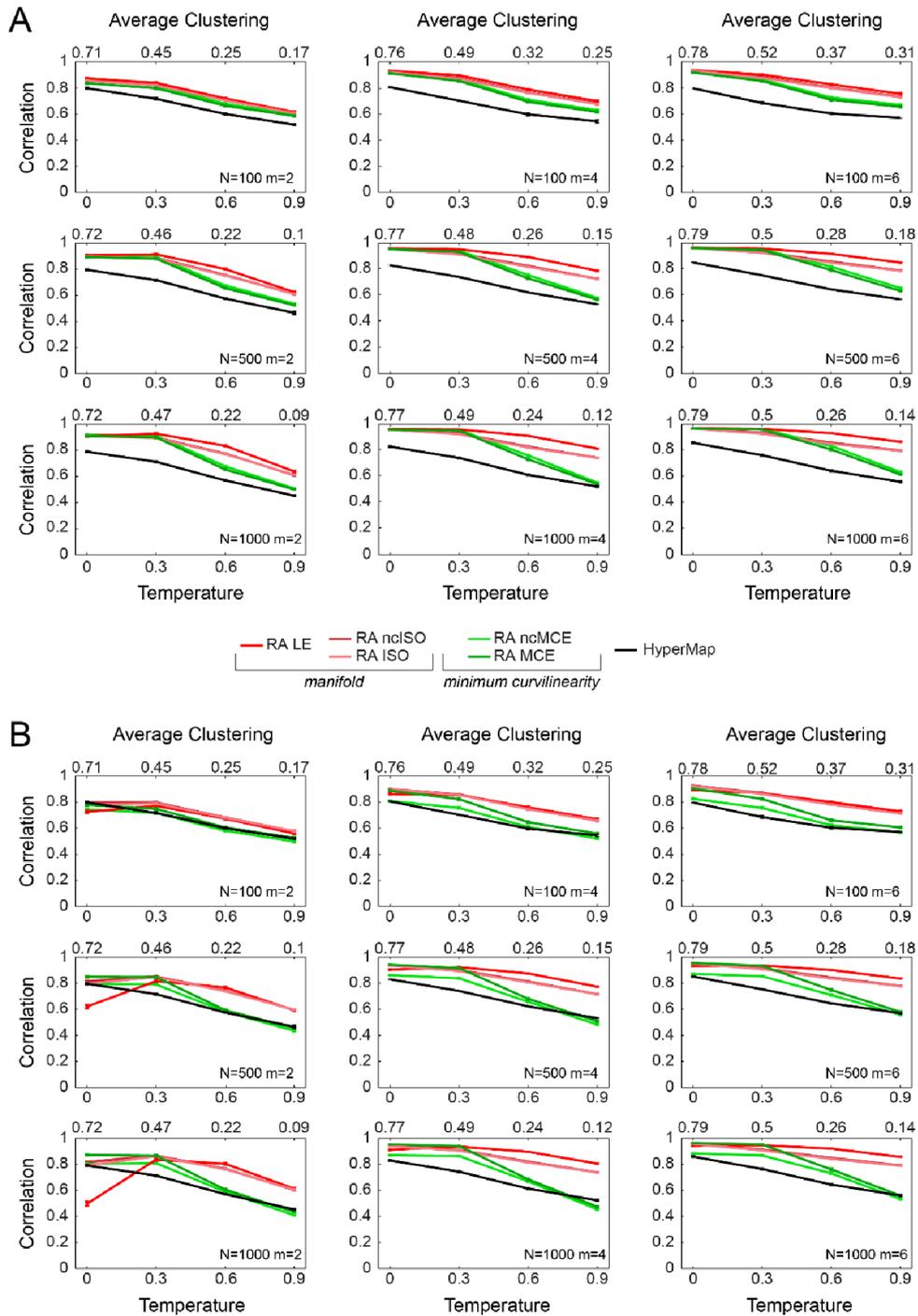

**Fig. 3. Evaluation on synthetic networks generated with the PSO model.** To validate the abovementioned techniques, we generated 100 different synthetic networks for each combination of tuneable parameters of the PSO model (temperature T, size N, average degree 2m, power-law degree distribution exponent γ). Suppl. Fig. 7 offers an idea of the topological diversity of the synthetic networks generated fixing γ = 2.5 and tuning the other parameters. In the results presented in the figures of this article we used γ = 2.5, but we also ran the simulations for γ = 2.25 and 2.75, and the differences were negligible (result not shown). The performance was evaluated as Pearson correlation between all the pairwise hyperbolic distances of the network nodes in the original PSO model and in the reconstructed hyperbolic space. The plots report the average correlation over the 100 synthetic networks that have been generated for each different parameter combination. The value one indicates a perfect correlation between the node hyperbolic distances in the original and reconstructed hyperbolic space. (**A**) Results of different methods when both RA and EA are applied. (**B**) Results of the different methods when RA is applied but EA is not. Comparing A and B, it is evident that the ability of EA to adjust for the local positional noise makes a difference. In fact, when m=2 and temperatures are low, RA-LE without equidistant adjustment (**B**) suffers a strong performance reduction in comparison to the case when EA is applied (**A**).

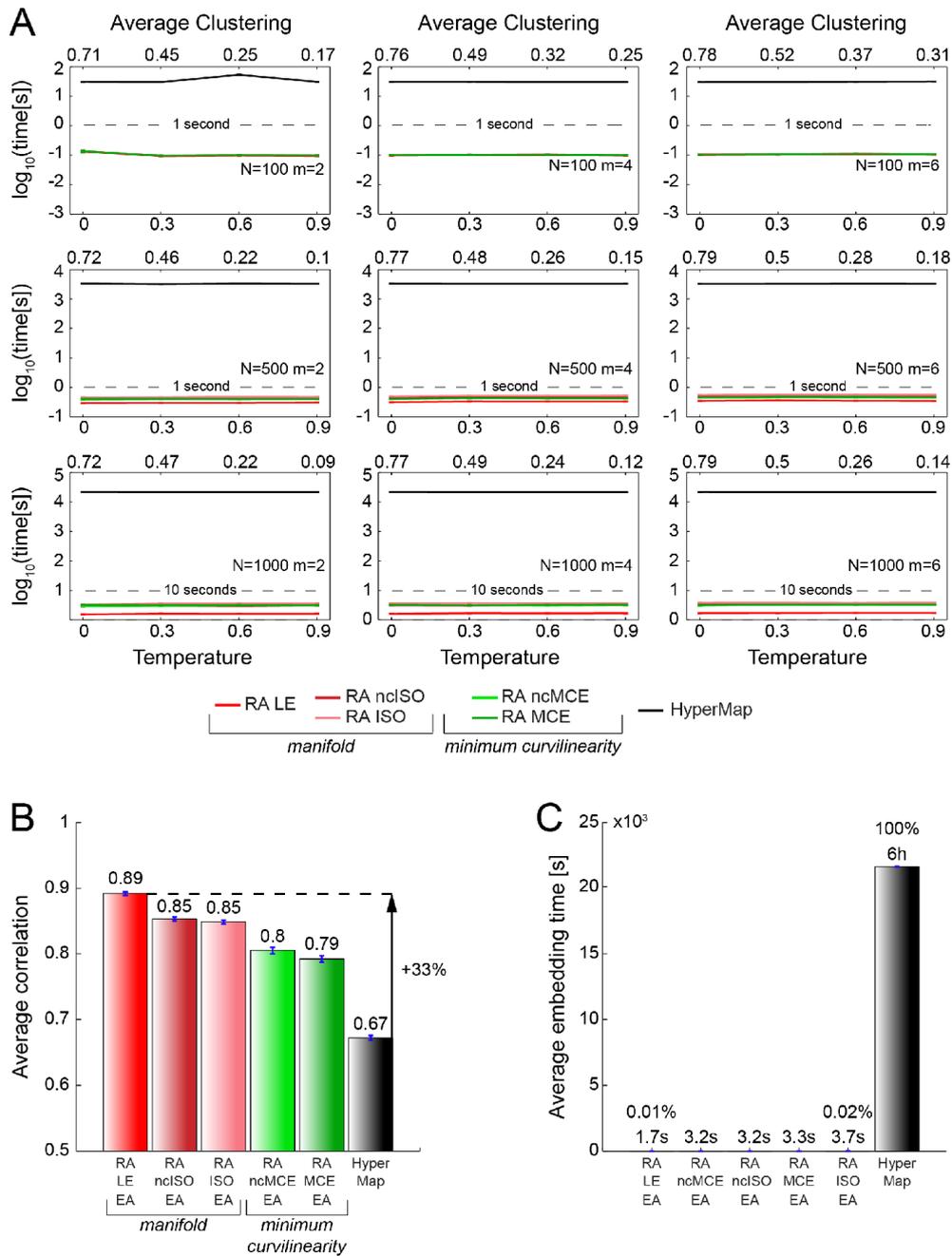

**Fig. 4. Time and performance comparisons. (A)** Computational time shows the large efficiency of the coalescent embedding based approaches that generally required less than 10 seconds to embed networks with 1000 nodes, while Hypermap spent approximatively 6 hours for the same task (software and hardware details in the methods). Finally, considering the average performance in all the simulations on 1000 nodes networks, coalescent embedding approaches achieved a performance improvement of more than 30% in comparison to Hypermap **(B)**, in the nearly 0.01% of the execution time **(C)**.

| Method | Karate N=34 E=78 T=0.41 γ=2.13 m=2.29 | Terrorist N=62 E=152 T=0.42 γ=3.87 m=2.45 | Dolphins N=62 E=159 T=0.70 γ=6.96 m=2.57 | Jazz N=198 E=2742 T=0.37 γ=4.48 m=13.85 | USAir N=332 E=2126 T=0.25 γ=1.82 m=6.40 | Flightmap N=456 E=37947 T=0.20 γ=1.71 m=83.22 | Median |
|---|---|---|---|---|---|---|---|
| RA_LE_**EA** | 8.44 | 12.42 | 4.04 | 6.69 | 62.64 | 8.22 | 8.33 |
| RA_MCE_**EA** | 8.60 | 10.81 | 3.42 | 6.10 | 59.57 | 8.03 | 8.31 |
| EBC_MCE_**EA** | 8.51 | 10.22 | 4.82 | 7.47 | 59.52 | 8.03 | 8.27 |
| EBC_ncISO_**EA** | 7.48 | 10.86 | 4.32 | 7.23 | 60.28 | 8.65 | 8.06 |
| RA_ncISO_**EA** | 7.79 | 12.31 | 4.05 | 6.71 | 62.38 | 8.26 | 8.03 |
| RA_ncMCE_**EA** | 7.65 | 10.53 | 2.91 | 7.73 | 62.37 | 8.31 | 8.02 |
| RA_ISO_**EA** | 7.65 | 11.80 | 3.99 | 6.72 | 59.18 | 8.25 | 7.95 |
| EBC_ncMCE_**EA** | 7.37 | 10.28 | 3.93 | 6.45 | 55.04 | 8.29 | 7.83 |
| LE_**EA** | 6.87 | 11.08 | 3.67 | 6.52 | 56.32 | 8.22 | 7.55 |
| EBC_LE_**EA** | 6.79 | 12.23 | 3.66 | 6.21 | 47.63 | 7.95 | 7.37 |
| EBC_ISO_**EA** | 5.92 | 11.20 | 4.13 | 6.10 | 56.12 | 7.91 | 7.01 |
| ncISO_**EA** | 6.52 | 6.97 | 4.04 | 6.30 | 56.93 | 9.05 | 6.75 |
| Hypermap | 7.94 | 7.97 | 2.42 | 5.10 | 46.22 | 4.67 | 6.52 |
| MCE_**EA** | 4.92 | 7.42 | 4.36 | 5.65 | 52.67 | 7.25 | 6.45 |
| ncMCE_**EA** | 5.32 | 8.69 | 4.29 | 5.65 | 52.57 | 7.18 | 6.41 |
| ISO_**EA** | 6.37 | 5.19 | 3.30 | 6.39 | 54.61 | 8.81 | 6.38 |
| EBC_MCE | 0.22 | 1.83 | 3.32 | 7.98 | 9.73 | 7.17 | 5.25 |
| EBC_ncISO | 1.22 | 2.91 | 3.12 | 7.21 | 12.22 | 8.65 | 5.17 |
| LE | 1.56 | 2.12 | 2.82 | 7.52 | 14.14 | 7.28 | 5.05 |
| EBC_ncMCE | 0.17 | 1.80 | 2.90 | 7.19 | 15.29 | 6.70 | 4.80 |
| EBC_ISO | 1.05 | 2.75 | 3.14 | 6.43 | 16.97 | 7.37 | 4.78 |
| RA_ncISO | 1.51 | 2.53 | 2.79 | 6.60 | 23.71 | 7.48 | 4.70 |
| RA_LE | 1.56 | 2.44 | 2.88 | 9.03 | 9.18 | 6.43 | 4.65 |
| RA_ISO | 1.46 | 2.39 | 2.73 | 6.36 | 19.58 | 7.56 | 4.54 |
| ncISO | 1.62 | 2.20 | 2.62 | 6.17 | 18.43 | 8.73 | 4.39 |
| RA_ncMCE | 0.10 | 1.78 | 1.95 | 7.71 | 16.69 | 6.83 | 4.39 |
| ISO | 1.43 | 2.05 | 2.27 | 6.26 | 15.05 | 8.39 | 4.27 |
| RA_MCE | 0.43 | 1.97 | 2.15 | 5.74 | 22.48 | 7.32 | 3.95 |
| EBC_LE | 1.21 | 2.22 | 2.57 | 5.86 | 2.67 | 3.24 | 2.62 |
| MCE | 0.52 | 0.39 | 2.58 | 6.45 | 4.19 | 1.03 | 1.81 |
| ncMCE | 0.52 | 0.50 | 2.56 | 6.52 | 3.80 | 1.03 | 1.79 |
| random | 1.00 | 1.00 | 1.00 | 1.00 | 1.00 | 1.00 | 1.00 |

**Table 1. Evaluation in link prediction on real networks.** Link prediction evaluation based on the method's precision with respect to a random predictor. One of the most accepted methods for link prediction evaluation was used (Lü et al. 2015; Cannistraci et al. 2013a; Daminelli et al. 2015). For each network 10% of the links is randomly removed and the remaining part of the network is embedded in the hyperbolic space. The hyperbolic distances between non-connected nodes are computed and ranked starting from the top in increasing order in a candidate list. A set of candidate links equal to the 10% removed is taken from the top of the candidate list and is compared against the set of real removed links. The proportion of correct links recovered represents the precision. This operation was repeated 500 times for each network and the mean precision for each method is reported. To compare the results across different networks relative precision with respect to a random predictor is reported for each network. The median performance on all the networks confirms that EA offers a strong boosting, and the methods at the first positions correspond with the best ones also in the evaluation on synthetic networks. Interestingly, the methods work fine also for networks with large gamma values and thus out of the typical scale-free range.

## Supplementary information

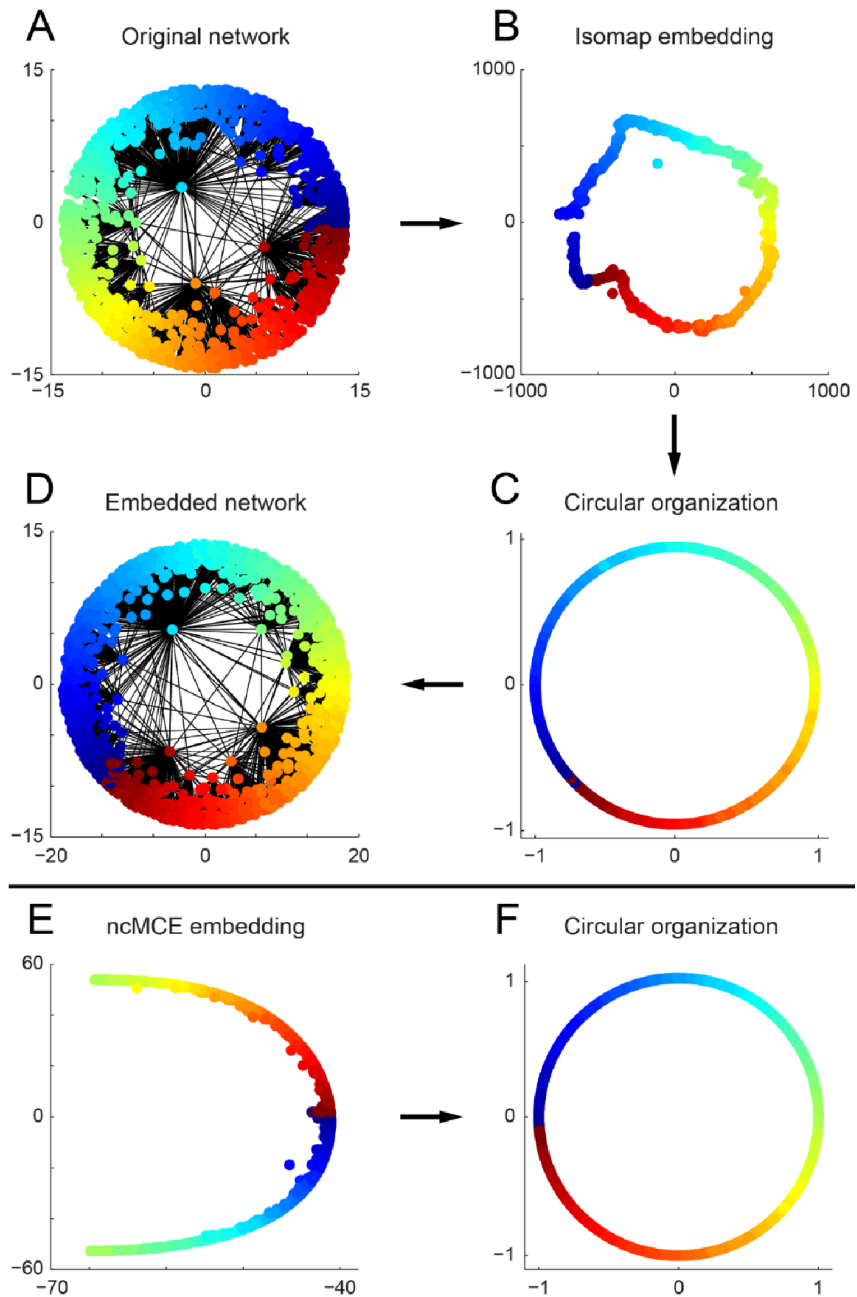

**Suppl. Fig. 1. Coalescent embedding with RA pre-weighting. (A)** We show the original synthetic network generated by the PSO model in the hyperbolic space. **(B)** The Isomap algorithm (ISO) starting from the adjacency matrix pre-weighted with the repulsion-attraction (RA) rule offers an embedding of the network nodes that is organized according to a circular pattern that follows the angular coordinates of the original PSO model. The circular pattern is visible more clearly compared to the embedding without the pre-weighting (Fig. 1). **(C)** The nodes are projected over a circumference and adjusted equidistantly according to the step 3.2 of the algorithm described in Fig. 2. **(D)** The radial coordinates are given according to the step 4 of the algorithm described in Fig 2. **(E)** A different pattern is obtained for ncMCE. The circular pattern is linearized and the nodes are ordered (respecting the node similarities) on the second dimension of embedding. **(F)** Accommodating the node points on the circumference following the ordering of the second dimension of embedding we can again recover an unbroken circular pattern that respects the angular coordinates of the original PSO model.

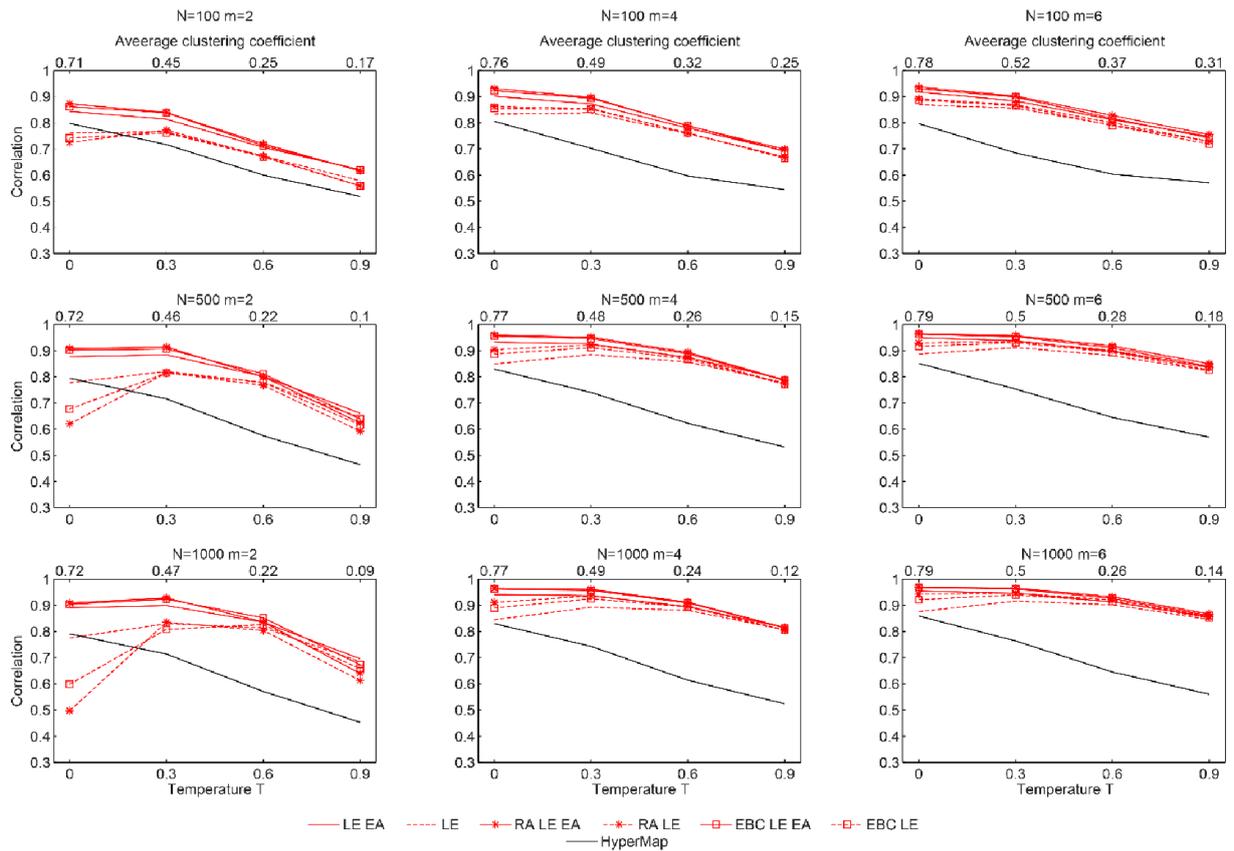

**Suppl. Fig. 2. Evaluation on synthetic networks: LE methods**

The figure shows the performance of all the methods using LE for dimension reduction. The performance was evaluated as Pearson correlation between all the pairwise hyperbolic distances of the network nodes in the original PSO model and in the reconstructed hyperbolic space. The plots report the average correlation over the 100 synthetic networks that have been generated for each different parameter combination as in Fig. 3.

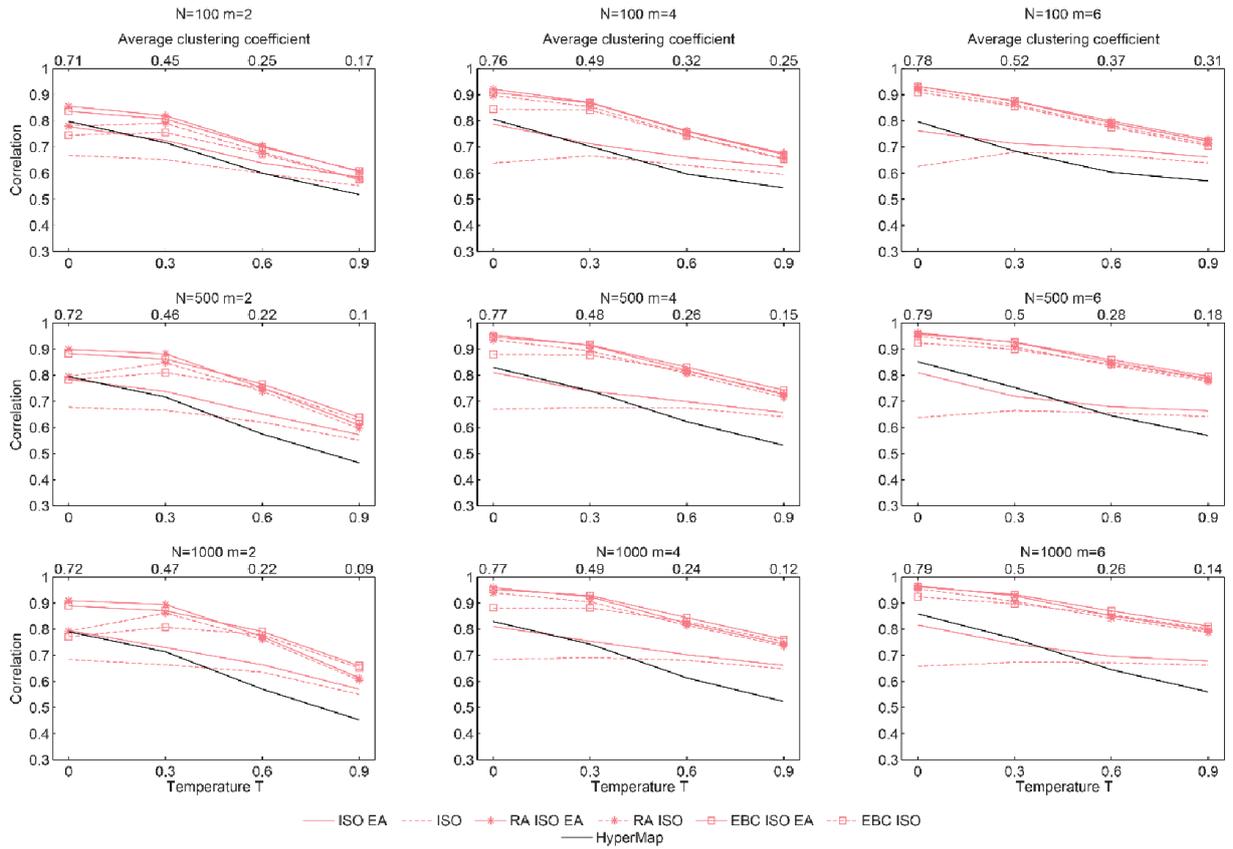

**Suppl. Fig. 3. Evaluation on synthetic networks: ISO methods**

The plots report the performance of all the methods using ISO for dimension reduction. The performance was evaluated as described in Suppl. Fig. 2.

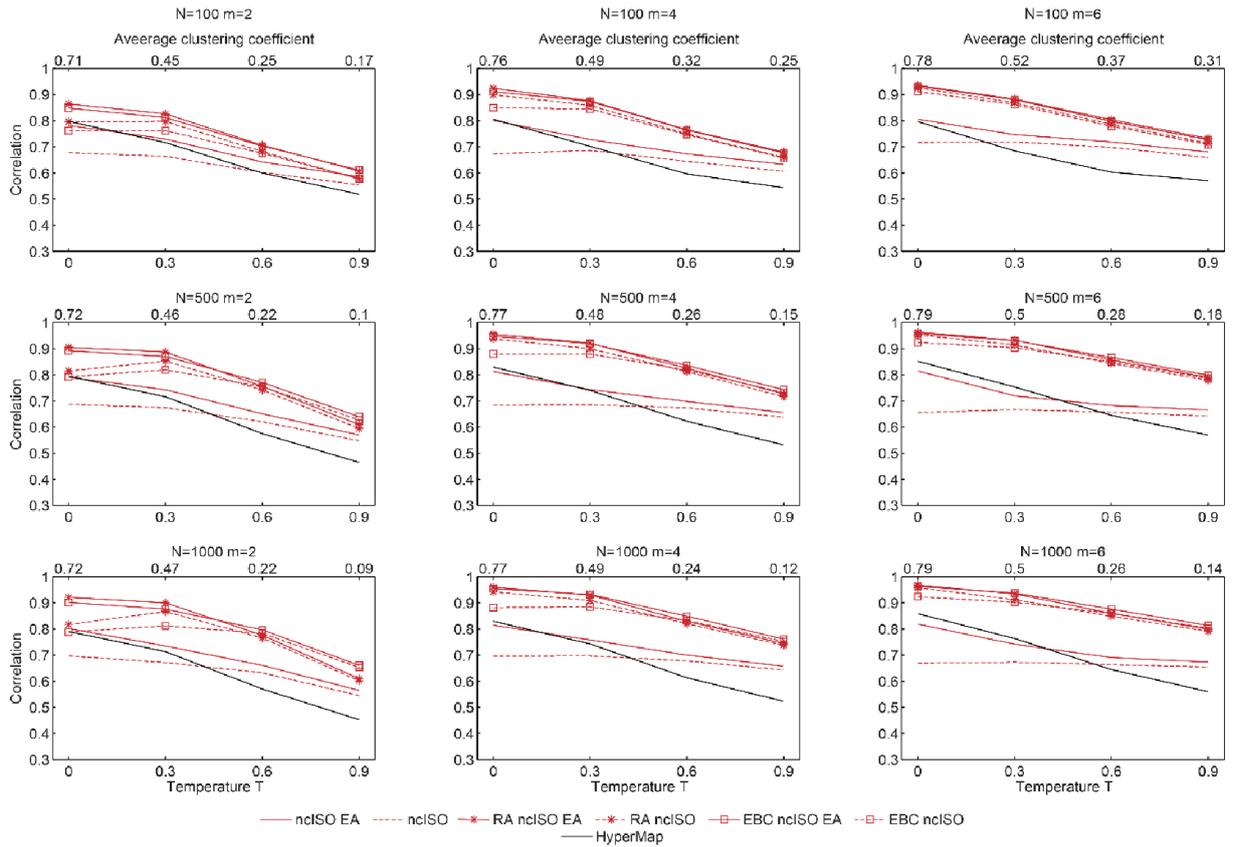

**Suppl. Fig. 4. Evaluation on synthetic networks: ncISO methods**

The plots report the performance of all the methods using ncISO for dimension reduction. The performance was evaluated as described in Suppl. Fig. 2.

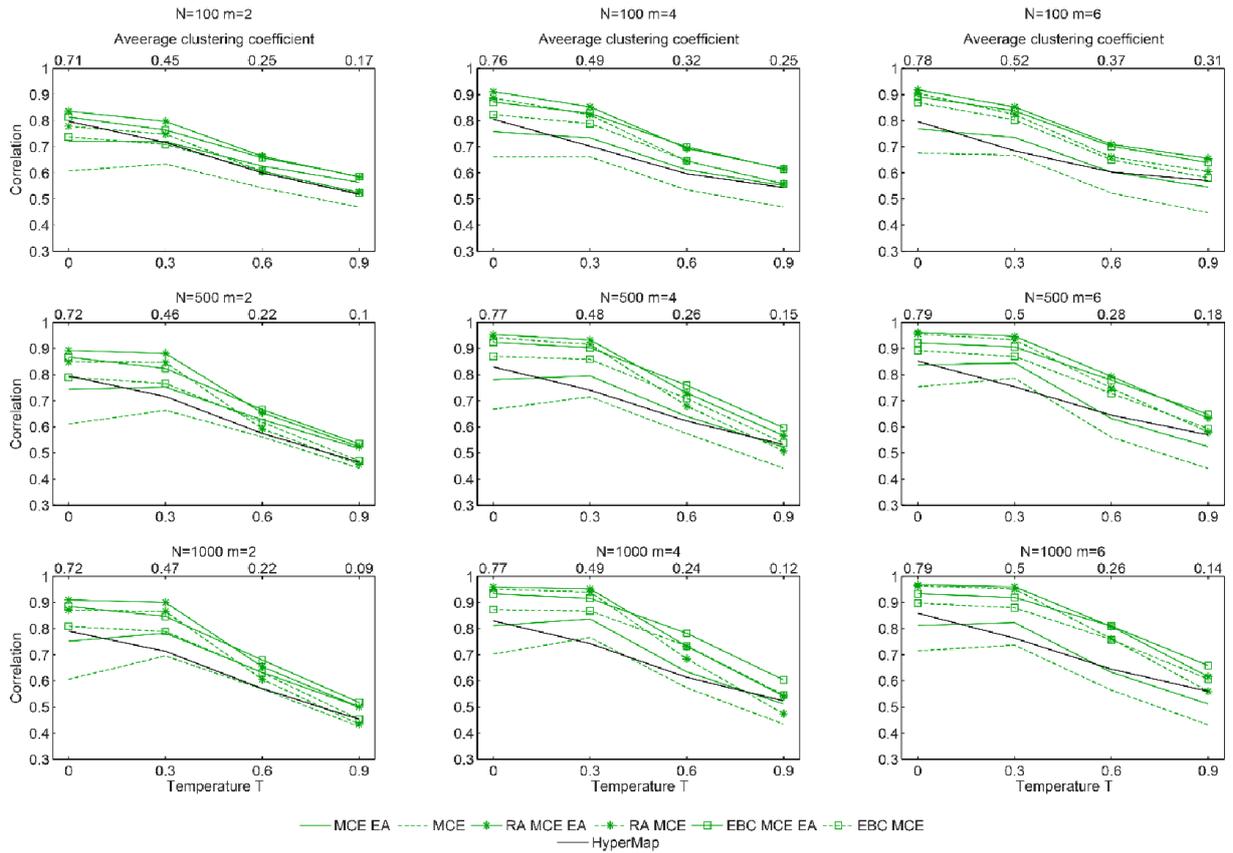

**Suppl. Fig. 5. Evaluation on synthetic networks: MCE methods**

The plots report the performance of all the methods using MCE for dimension reduction. The performance was evaluated as described in Suppl. Fig. 2.

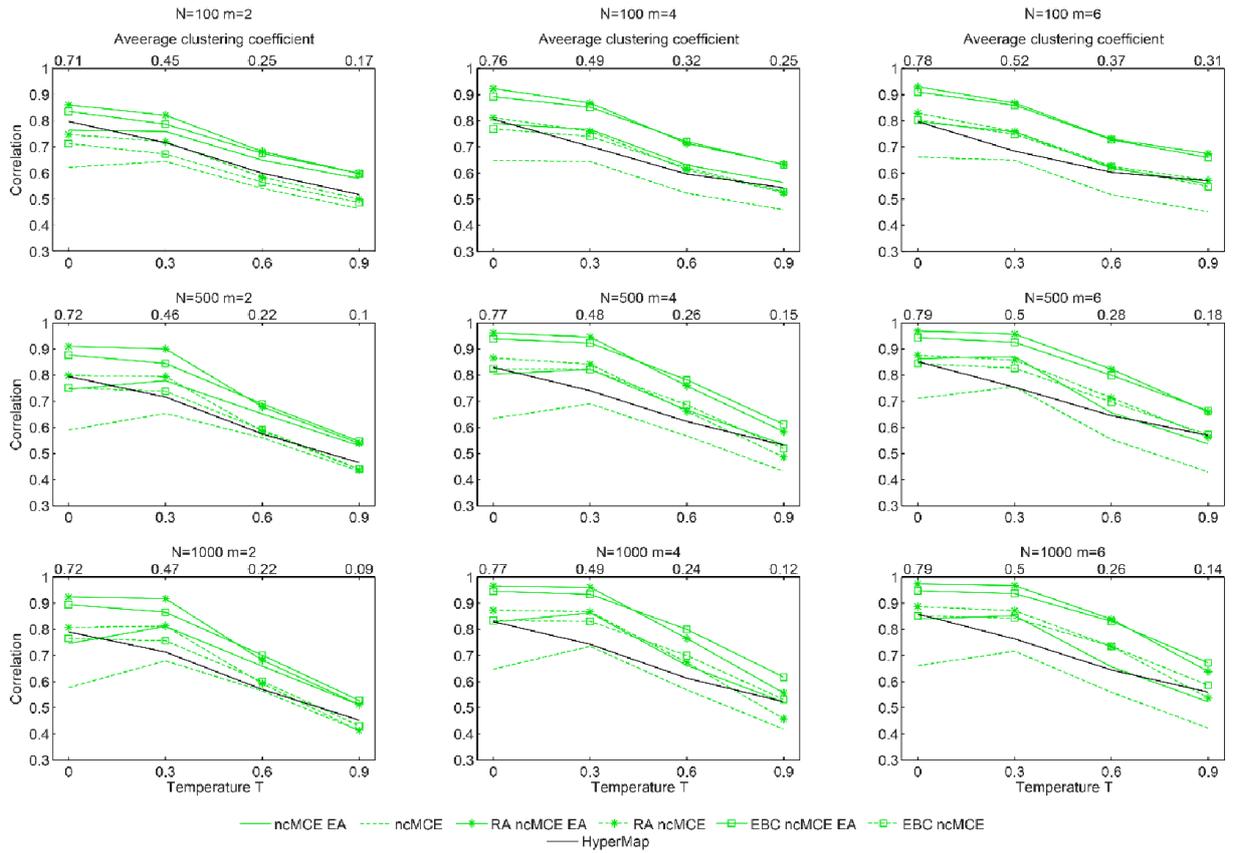

**Suppl. Fig. 6. Evaluation on synthetic networks: ncMCE methods**

The plots report the performance of all the methods using ncMCE for dimension reduction. The performance was evaluated as described in Suppl. Fig. 2.

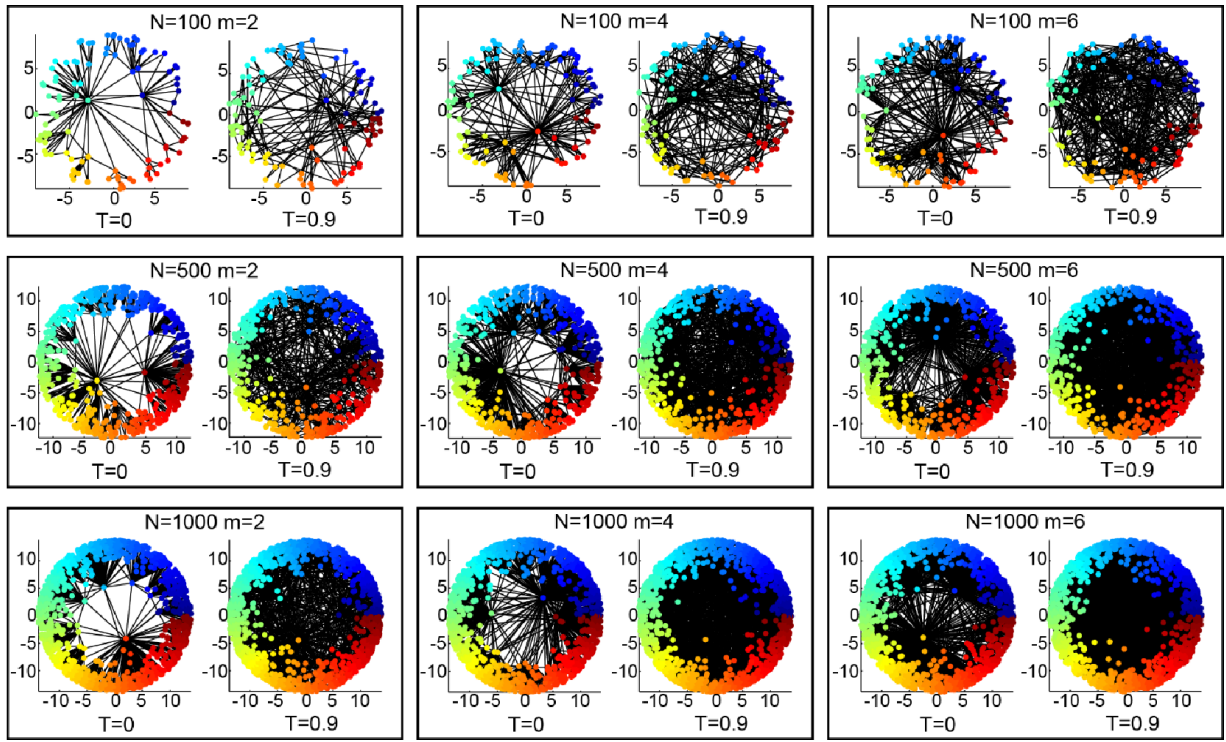

**Suppl. Fig. 7. Examples of synthetic networks generated with the PSO model**

Examples of synthetic networks for each parameters combination of N and m, at T = 0 and T=0.9. This illustrates the increase of randomness and the decrease of clustering at high temperatures.

**Real network datasets**

Table 1 shows the results of different methods for 7 real social interaction networks, which represent differing systems: (1) karate, (2) terrorist, (3) dolphins, (4) jazz, (5) USAir, (6) flight map.

All the networks are undirected and unweighted and were ordered in the table in ascending order with respect to the number of nodes (N), taking also into account the number of edges (E) when the same number of nodes is present.

The first network is of Zachary's Karate Club, it represents the friendship between the members of a university karate club in US; 34 nodes and 78 edges (Zachary, 1977).

The second network is a terrorist association network, responsible for the 9/11 attacks; 62 nodes and 152 edges (Krebs, 2002).

The third network is a social network of frequent associations between (bottlenose) dolphins in a community living off Doubtful Sound, New Zealand; 62 nodes and 159 edges (Lusseau & Newman, 2004).

The fourth network is a collaboration network of jazz musicians; 198 nodes and 2742 edges (GLEISER & DANON, 2003).

The fifth network represents the USAir transportation network; 332 nodes and 2126 edges (Batagelj & Mrvar, 2006).

The sixth network depicts the flight map between pairs of American and Canadian cities; 456 nodes and 37947 edges (Frey & Dueck, 2007).